# Why do Particle Clouds Generate Electric Charges?

T. Pähtz[1], HJ Herrmann[1] & T. Shinbrot[1,2]
[1] Institute für Baustoffe, ETH-Zürich, CH-8093 Zürich, Switzerland
[2]Department of Biomedical Engineering, Rutgers University, Piscataway, NJ 08854, USA

Grains in desert sandstorms spontaneously generate strong electrical charges; likewise volcanic dust plumes produce spectacular lightning displays. Charged particle clouds also cause devastating explosions in food, drug, and coal processing industries. Despite the wide-ranging importance of granular charging in both nature and industry, even the simplest aspects of its causes remain elusive, for it is difficult to understand how inert grains in contact with little more than other inert grains can generate the large charges observed. In this paper, we present a simple yet predictive explanation for the charging of granular materials in collisional flows. We argue from very basic considerations that charge transfer can be expected in collisions of identical dielectric grains, and we confirm the model's predictions using discrete element simulations and a tabletop granular experiment.

## Introduction

As long ago as 1850, Michael Faraday commented on the peculiarities of the production and discharge of electric charges during sandstorms[1], a phenomenon repeatedly rediscovered over the intervening century and a half[2,3,4,5,6,7]. Similarly, sand is known to become strongly electrified by helicopters traveling in desert environments, producing spark and explosion hazards[8], and the issue even has implications for missions to the Moon and to Mars[9,10], where charged dust degrades solar cell viability and clings to spacesuits, limiting the lifetime of their joints[11]. Several research groups have investigated mechanisms by which similar particles may charge one another, for example due to non-uniform heating[12], differences in contact area[13] or particle size[14,15], inductive charging of isolated particles[16], or aqueous ion transfer at particle surfaces[17]. Recent work has also revealed that identical water droplets can acquire and transfer net charge at minute points of contact[18].

Notwithstanding these developments, the phenomenon of granular charging remains poorly understood for want of adequate explanations for two very basic and well-documented facts. First, insulators – which by definition have no free charge carriers to perform the task – transfer large amounts of charge[8,12,19], and second, identical materials – such as grains of sand in the desert – are known to charge one another on contact[12,13,15,20]. In the present article, we propose a mechanism to address these twin conundrums. To do so, we note that granular charging predominates for insulating materials under dry conditions, and indeed, first hand reports state that charges in sandstorms dissipate rapidly on the onset of rain[21]. Under such insulating conditions, charge should not be transported either by the insulating grains or by the dry and insulating environment; on the contrary, charged insulators should be expected to neutralize at points of contact. We therefore propose a mechanism by which neutralization of particles near their points of contact can generate the seemingly paradoxical increase[15] in granular charges.

We begin by considering a caricature of a collision of two grains within a strong electric field – as is documented to be ubiquitous within charged dust clouds[22,23,24]. As shown in Fig. 1, if the grains are initially electrically neutral and both grains and their environment are sufficiently insulating, the effect of an electric field, **E**, will be to polarize the grains. We depict this in Fig. 1 as producing negatively charged upper and positively charged lower hemispheres. The simplest case occurs when the grains collide and respective hemispheres become neutralized, as indicated in the center panel of Fig. 1. Real collisions, between non-spherical particles containing complex charge distributions[25], would certainly be much more complicated than this cartoon can capture, however this simplified model has the merits that it can be fully analyzed, and as we will show, it provides experimentally testable predictions.

The result of the caricatured collision shown in Fig. 1 is that the top- and bottom- most hemispheres of the granular assembly retain a charge, while the contacting hemispheres become neutralized. After the collision, as sketched in the right panel of Fig. 1, each individual grain is again exposed to the pre-existing electric field, causing the grains to be repolarized with additional unit charges top and bottom. As the right panel indicates, the result of this process is to increase the negative charge by one unit on the upper particle, and the positive charge on the lower one by the same amount.

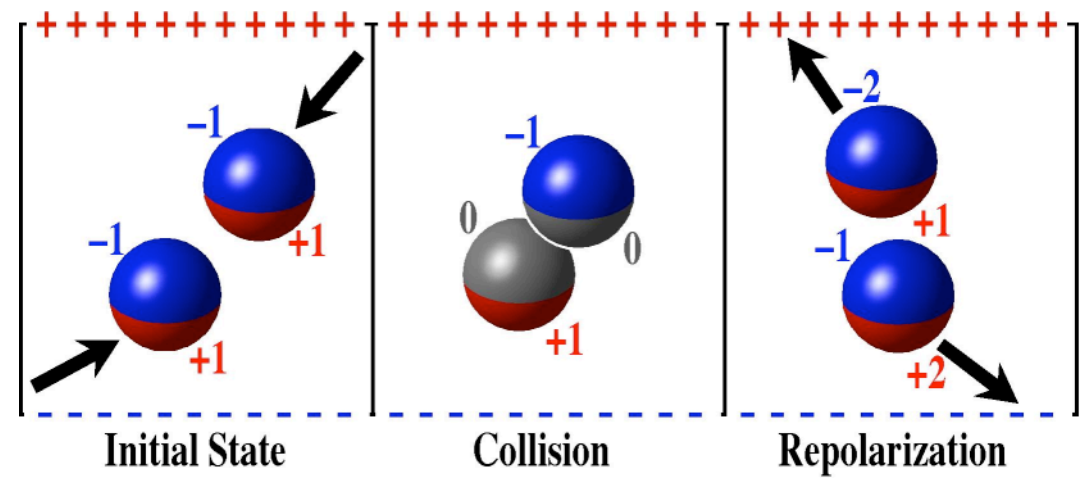

*Figure 1 – Proposed mechanism of charging of colliding granular particles in an electric field. Initially (left panel) a pair of polarized particles with two units of opposite charge produced by the electric field collide (center panel) to neutralize adjoining hemispheres, and once separated (right panel), the particles again become polarized by the external field. In this way, a pair of initially neutral but polarized particles becomes charged, in this case with the upper particle more electro-negative and the lower particle more positive. In this simplified model, provided that the electric field generates a constant polarization and that collisions act to neutralize the top or bottom hemisphere of each particle, particles gain one unit of charge following every collision. In this figure, blue denotes negative and red positive charge as indicated by the numbers beside each hemisphere, and the arrows indicate representative particle velocities.*

This charge transfer occurs for every collision, and so in this scenario, collisional granular flows should pump positive charges downward to ground and negative charges upward to the top of an agitated bed at a predictable rate, proportional to the collision frequency in the dust cloud. We can therefore estimate the rate of particle charging for this model from kinetic theory, which provides that the collision rate, R, for moving particles is simply[26]:

$$R = C n^2 V_{rms}. \quad [1]$$

Here C is a constant proportional to the particle cross-section, n is the number density of particles, $V_{rms}$ the mean particle velocity, and R has units of collisions per unit time. For granular flows, $V_{rms}$ at steady state is achieved by a balance between the rate of energy input and the rate of dissipation. We consider here the situation in which energy is input from below - as is documented to occur when windblown particles strike the ground[27] - and in which dissipation occurs during inelastic particle collisions.

To make the problem analytically tractable, we specify that whenever a grain strikes the ground, it is ejected upward with velocity, $V_o$ that is diminished by a fixed restitution coefficient, ε, by each overlying layer of particles that the grain passes through. We assume that the velocity is diminished up to a maximum number of layers, $L_{max}$, beyond which no further impulse is transmitted. In this case, we can write that:

$$V_{rms}(L) = \sqrt{\frac{\int_1^L \left(V_o \varepsilon^{\ell}\right)^2 d\ell - \nu^2}{L}}, \quad [2]$$

where L is the number of layers of particles through which an ejected particle may pass, and ν is a constant that ensures that $V_{rms} = 0$ when $L = L_{max}$. If we make the first order approximation (which we validate with computations shortly) that the particle density grows linearly with L, then after insertion of Eq. [2] into Eq. [1], we obtain:

$$R = \alpha L^{3/2} \sqrt{\varepsilon^{2L} - \beta}, \quad [3]$$

where the constant $\alpha = CV_o \big/ 2\sqrt{|\ln \varepsilon|}$ is determined by the particle cross-section C, the velocity $V_o$, and the coefficient of restitution ε, while the constant β depends on gravity and ε. In practical terms, Eq. [3] predicts that the charging rate should be small for both very shallow and very deep agitated beds: for shallow beds (i.e. small L), the charging rate will be small because the number density will be small and hence particle collisions will be infrequent, whereas for deep beds (large L), the charging rate will be small because collisions will be numerous, and so the finite coefficient of restitution will cause the bed to collapse. We remark that the rapidity of the dropoff in charge at large L is regulated by the parameter β: for large β the dropoff is abrupt; for smaller β, the dropoff is more gradual, and that the dimensional charging rate is R multiplied by the unit charge imparted per collision.

To test this model, we perform simulations and experiments in which inelastic particles are agitated from below and we evaluate the accumulation of charge in the presence of an externally applied electric field.

## Simulation

The simulation we use is modeled after Walton and Braun[28], and tracks the motions of polydisperse spherical particles that collide inelastically in three dimensions under the influence of gravity.

After each collision, the net charge on each particle is recalculated and a vertical force is applied that is proportional to the product of that charge and an external, vertically oriented, electric field of fixed strength. As described in Fig. 1, charges on upper and lower hemispheres of each particle neutralize during every collision, and each particle is repolarized by adding opposite unit charges to its top and bottom following the collision. To keep upper and lower charged hemispheres aligned vertically, collisions are taken to be frictionless though inelastic with coefficient of restitution 0.94 (a value that generates a fluidized bed similar to that used in comparison experiments discussed shortly). Whenever a particle strikes the bottom of the simulated volume, both hemispheres of the particle are neutralized, and to mimic the so-called 'splash' that particles impacting on a sand bed produce during Aeolian transport[27], the particle is ejected vertically with velocity $V_o = 2.7\sqrt{gd}$, a value that empirically produces granular fluidization over a wide range of parameter values.

The simulated volume is periodic in the horizontal directions. We have also performed simulations using fixed walls, however these results do not differ noticeably from those shown here and we omit them from our discussions. Likewise separate simulations using horizontal dimensions of 8×8, 11×11, and 13×13 mean particle diameters yield indistinguishable results provided that the depth of the bed (discussed shortly) is held fixed, so the data shown are for 8X8 diameter periodic domains. Particles that acquire charge greater than m·g/E, where m is the particle mass, g is gravity and E is the applied electric field strength, are removed from the simulation once they are out of contact with all other particles. These particles are then replaced by particles of zero charge beneath the simulation with upward speed $V_0$. The nominal depth of the bed is counted in number of layers, L, where one layer consists of the number of particles (about 62) that can be placed in a monolayer in this domain. We assume that this nominal depth defines the number of layers of particles through which a grain ejected at the bottom of the bed must pass (L in Eq. [2]).

Typical results of simulations are shown in Fig. 2(a) for 288 particles (4.6 layers), 540 particles (8.7 layers), and 828 particles (13.4 layers). Particles are coded depending on their net charge as defined in the color bar. The mean charge per particle reaches a steady asymptotic state within about $5\cdot 10^6$ time iterations: we have extended simulations in several representative cases out to $10^9$ iterations, and we find no detectable differences in the spatial distributions of particles or their charges. Qualitatively, it is apparent from Fig. 2(a) that the most strongly charged particles are near the top of the bed, and that many more particles are highly charged (red) for intermediate numbers of layers than for either high or low numbers, as predicted by Eq. [3]. A quantitative comparison between the mean charge per particle from the simulation (blue solid line) and the fit predicted from Eq. [3] using $\beta$=1 is shown as a dashed line in Fig. 2(b), which confirms the qualitative impression of Fig. 2(a).

As we have described, the derivation of Eq. [3] depends on two essential assumptions. First, the number density is assumed to grow approximately linearly with L, which implies that the depth of the agitated bed should not grow as L is increased. This is what leads to low density and hence weak charging at small L. We remark that there is no analytic framework to predict how the depth of an agitated and charged bed will depend on L, however our computational results shown in Fig. 2(a) indicate qualitatively that the depth of the agitated bed remains comparable as L is increased. This can be quantitatively confirmed by evaluating the number density, n, of the bed as a function of height. This is plotted in Fig. 2(c), where we have calculated n by dividing the computational volume into horizontal slices and counting the number of particles within each slice. Consistent with Eq. [3], grains apparently extend to about 13 or 14 mean grain diameters irrespective of L, thus the volume of the agitated bed does not depend significantly on L and it appears justifiable to set the particle density proportional to L. The second assumption underlying Eq. [3] is that $V_{rms}$ diminishes with L due to inelasticity of particle collisions. This is what produces weak charging at large L. Fig. 2(c) shows that beds 11 layers or deeper attain a solidified, nearly random close packed, state, while shallower beds maintain a number density below 50%. Thus as predicted by Eq. [3], our simulations confirm that shallow beds remain fluid-like and so charge weakly because their number density diminishes with L, while deep beds solidify and so charge weakly because their collisional velocities are suppressed.

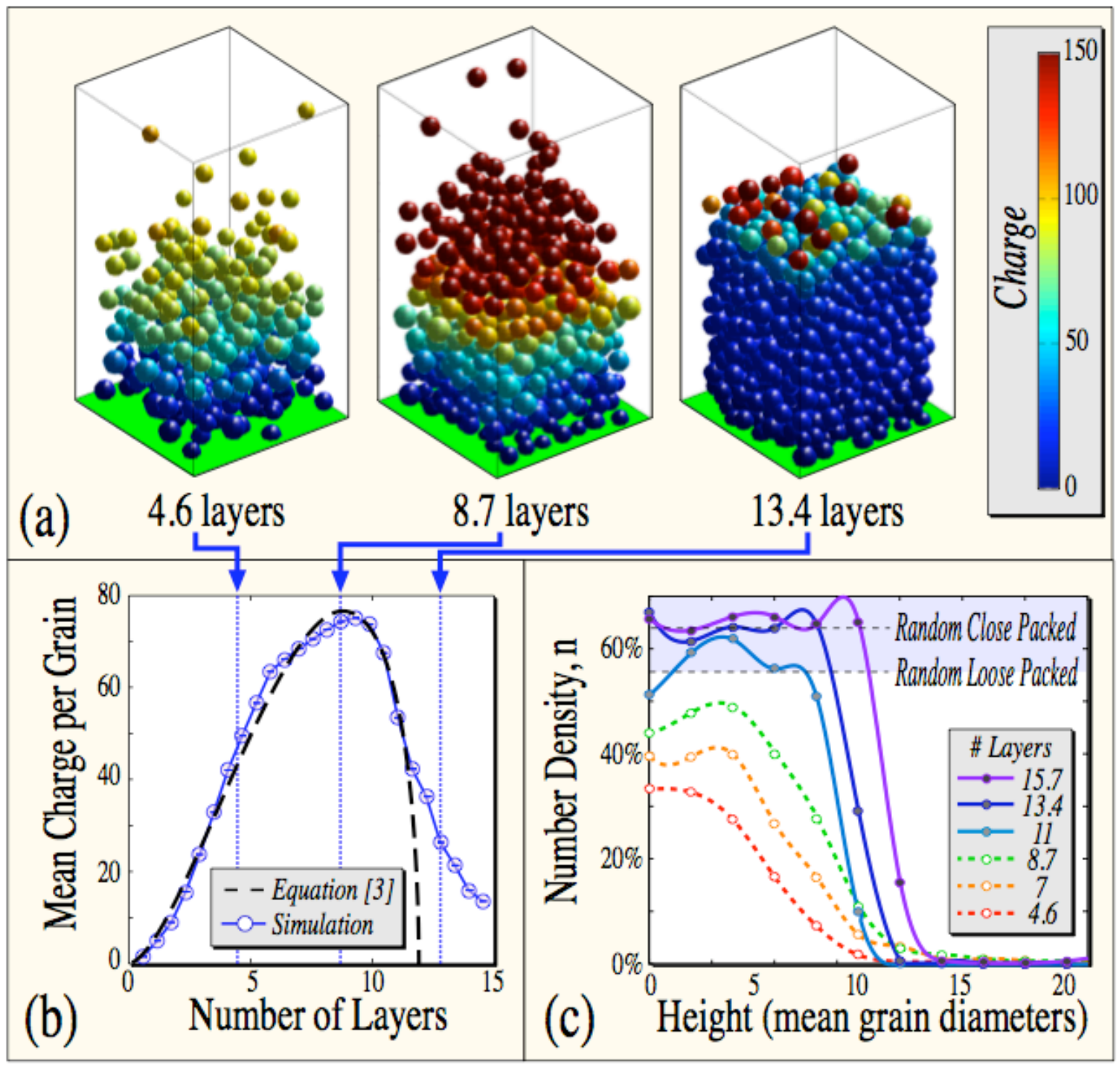

*Figure 2 – Simulation results. (a) Snapshots of granular bed at three representative depths of the granular bed after $2 \cdot 10^7$ computational timesteps, color coded by particle charge. This panel shows that qualitatively, shallow and deep beds produce less granular charge than intermediate depth beds. (b) Quantitative evaluation of mean charge per grain versus bed depth. Each data point is an average of 750 measurements of mean bed charge taken at regular intervals between $5*10^6$ and $20*10^6$ iterations. Error bars are included, but are smaller than the plot symbols. The dotted curve is a fit to Eq. [3] using coefficient of restitution $\varepsilon = 0.7$, about that of glass. (c) Evaluation of the mean solids fraction versus height, fit with cubic splines (error bars over replicate simulations are again smaller than the plot symbols). The lower parts of beds deeper than 11 layers (solid curves) attain a solidified state between random loose and random close*

*packed densities shown. Shallower beds (dotted curves) are nowhere solidified, thus grains remain in motion everywhere.*

## Experiment

Our model and simulation hinge on simplifications whose validity remains to be demonstrated. To test our theoretical and computational results, we constructed a "spouted bed," in which colored glass beads of mean diameter 1.6 mm are fluidized by air blown from below through a porous plenum 6 cm in diameter. As shown in Fig. 3(a), the experiment is contained in a 5 mm thick glass jar about 25 cm diameter at its base, that is separated by a small distance, to allow for air egress, from a grounded metal supporting plate. In each experiment, the airflow is set to the lowest pressure at which the grains above the plenum just become fluidized, so that by design grains charge only by contact with one another or with the grounded plenum. An external electric field is applied by placing a second metal plate that is connected to a 30 kV van de Graaff generator above the apparatus and outside of the jar. As shown in the enlargements in the lower panels of Fig. 3, shallow beds only weakly fluidize, while deeper beds become energetically agitated. In both cases, grains float spontaneously within the chamber and hover or bounce against the upper surface. Movies are included in supplemental materials (http://coewww.rutgers.edu/~shinbrot/Sandstorms). When the generator is turned off, grains remain adhered both to the top surface and to the side of the glass jar (Fig. 3(e)). We emphasize that since the upper plate was at a high positive potential, only negatively charged grains could remain adhered to the nearby glass, yet the bottom plate is grounded, and there is no source of negative charge anywhere within the glass jar.

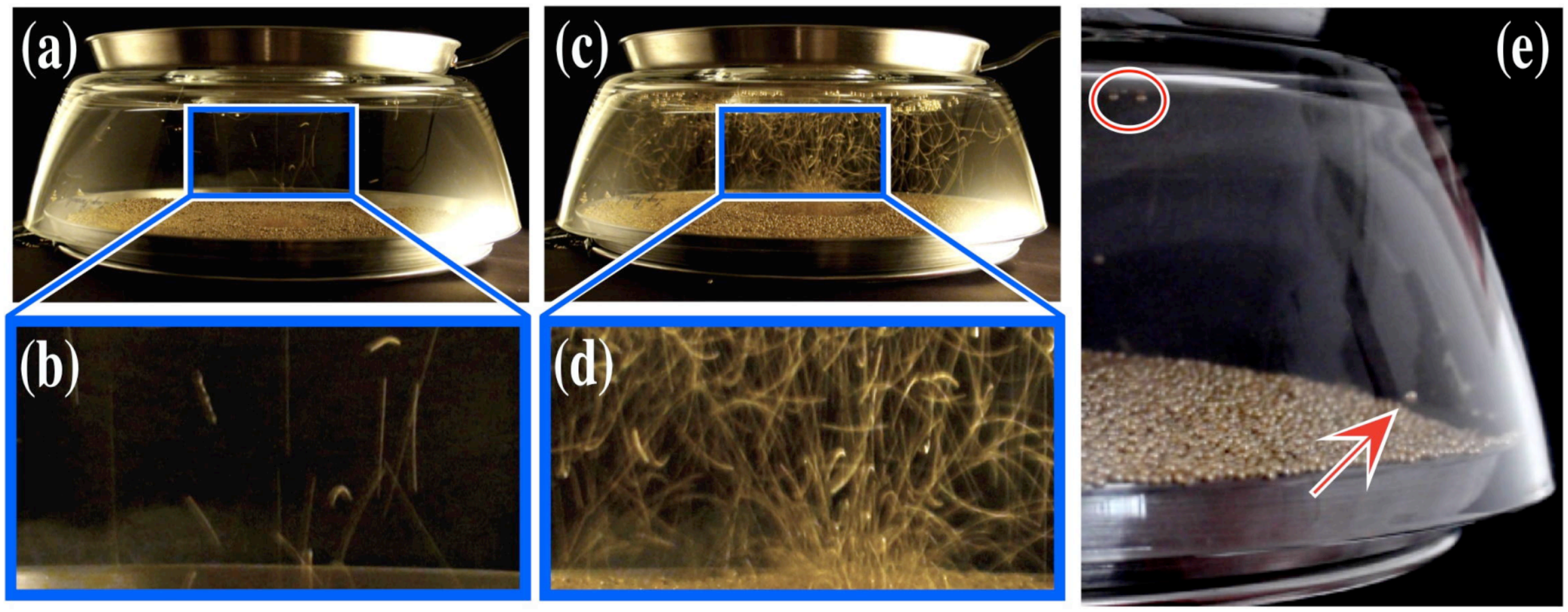

*Figure 3 –Views of experiment for (a) 4 and (c) 10 layers of particles. Upper panels show the spouted bed, fed from below by an airstream through a porous plenum at the center of the bottom plate. Lower panels show enlargements of upper window of the agitated bed. The plenum and the plate are conductive and grounded, the metal plate above the glass jar is connected to a 30 kV voltage generator, and the container itself (about 30 cm in diameter) is sealed except for a < 1 mm gap around the edge of the bottom plate to allow air to escape. Some grains adhere to the top (circle) or sides (arrow) of the container, as shown in (e) after the air flow has been halted. Particles are colored glass beads of diameter 1.6 ± 0.1 mm, and the RH is measured by sling psychrometer to be 51±2%.*

In this experiment, measuring actual particle charge is problematic since particles are deliberately isolated inside a glass enclosure and the entire experiment is exposed to a strong electric field that would interfere with any sensitive charge measurement. As a surrogate for particle charge, we evaluate the number of levitated particles within a fixed window between the granular bed and the top of the glass jar. This number is manually counted in 20 successive snapshots taken at 2 second intervals, and the resulting average is plotted in Fig. 4 as a function of the number of layers, L, of grains. To obtain this plot, L is determined by weighing the number of grains that will fit in a monolayer on the bottom plate of the experiment. Successive multiples of this weight of grains are then loaded and leveled in the apparatus, so that one monolayer gives L = 1, two monolayers give L = 2, etc. This is the identical procedure used to define numbers of layers in the simulations that we have discussed. Once a fixed number of layers is loaded into the apparatus, the pressure is then adjusted, as we have described, to the minimum value at which grains remain fluidized, and then the van de Graaff generator is turned on, snapshots are taken, and numbers of levitated grains are counted and averaged.

In the inset to Fig. 4, we show for comparison the number of levitated grains from the simulations previously described. For the simulation, we count the number of particles in a fixed size window between 15 and 21 mean particle diameters from the bottom of the bed - a distance above any solidified substrate as shown in Fig. 2(c). In both the main plot and the inset, we overlay the dashed curve from Fig. 2(b), rescaled and offset to account for the fact that charges below a fixed threshold cannot be expected to levitate finite weight particles.

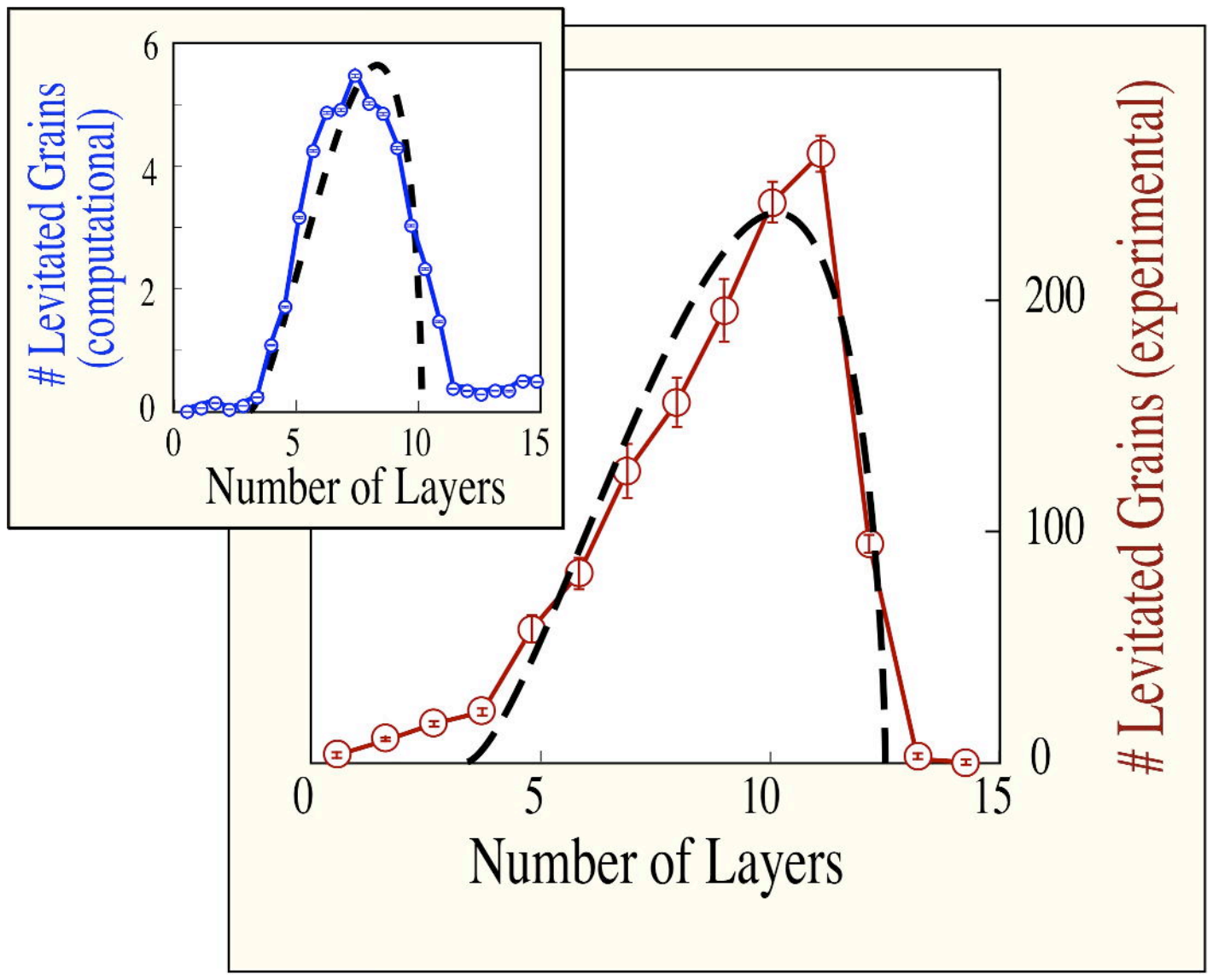

*Figure 4 – Numbers of levitated grains in experiments (main plot) and simulations (inset). The computational number of grains is an average over 9500 independent realizations; the experimental number is summed over 20 successive snapshots taken at 2 second intervals. The experiments were performed over the course of several days, with RH ranging between 45% and 53% ± 2%. The dashed lines are plots of Eq. [3] from Fig. 2(b), rescaled and offset by 3 layers as described in text.*

## Conclusion

We have introduced a simplified model that appears to accurately predict the charging of granular materials in collisional flows such as those encountered in particle clouds such as sandstorms, volcanic plumes, or industrial fluidized beds. Our simulations and experiment confirm the essential features of the model, namely that identical grains in the presence of an applied electric field can pump charge upward through repeated collisions in the absence of any conductive mechanism of charge transfer either in the particles or their environment. We find as predicted that shallow agitated beds - as could be expected in weak winds or for heavy grains - charge weakly, as do very deep agitated beds - as would be expected for highly dissipative materials. Under intermediate conditions, however, we observe dramatic charging, with the most highly charged particles found preferentially near the top of the agitated bed.

We emphasize that this charging mechanism has nothing to do with electrochemical differences in surface states, or variations in sizes or types of contact. Such differences do unequivocally lead to charging, however not for identical materials under consideration in this study. All that is needed is repeated collisions between dielectric particles in the presence of a sufficiently strong electric field. The charging effect reported here appears to be robust: indeed, the experiments shown were performed at moderately humidity, between 45% and 53% RH, but similar effects have also been seen in our laboratory at RH down to about 20%. Moreover, despite the fact that we used beads large enough to facilitate counting in the experiments described, we have reproduced the vigorous charging and levitation of grains using smaller and irregular particles as well.

In closing, we stress that although this work explains how grains in an electric field can acquire strong charges, it does not define mechanisms that may generate the required electric field. Such fields are well documented to exist[22,23,24], yet their cause in natural sandstorms is poorly understood. In some cases, the source may be external, as in reports that nearby thunderstorms[5] or charged bodies[29] can provoke granular charging. In other cases, it remains to be determined how a sandstorm might both generate strong charges and produce the electric field that engenders the charging to begin with. We hope that larger scale studies can both probe the accuracy of the simple model presented here and identify mechanisms by which a self-sustaining electric fields may be established.

Acknowledgments: We thank E. Strombom for her dedicated experimental work, and we thank the National Science Foundation, Division of Chemical and Transport Systems and the Eidgenössische Technische Hochschule, project ETH-10 09-2 for financial support.